\documentclass[aps,pre,twocolumn,groupedaddress,showpacs,floatfix]{revtex4}
\usepackage{graphicx}
\usepackage{pst-all}
\usepackage{color}
\usepackage{amsmath}
\usepackage{textcomp}
\newcommand{\comment}[1]{}

\begin{document}
\title{Relevant components in critical random Boolean networks}
\author{Viktor Kaufman and Barbara Drossel}
\affiliation{Institut f\"ur Festk\"orperphysik,  TU Darmstadt,
Hochschulstra\ss e 6, 64289 Darmstadt, Germany }
\date{\today}
\begin{abstract}
Random Boolean networks were introduced in 1969 by Kauffman
as a model for gene regulation.
 By combining analytical arguments and efficient numerical simulations,
we evaluate the properties of relevant components of critical random
Boolean networks independently of update scheme.
 As known from previous work, the number of relevant
components grows logarithmically with network size. We find that in
most networks all relevant nodes with more than one relevant input sit
in the same component, while all other relevant components are simple
loops. As the proportion of nonfrozen nodes with two relevant inputs increases, the number of relevant components decreases
and the size and complexity of the largest complex component grows.
We evaluate the probability distribution of different types of
complex components in an ensemble of networks and confirm that it
becomes independent of network size in the limit of large network size.
In this limit, we determine analytically the frequencies of occurence 
of complex components with different topologies.
\end{abstract}
\pacs{89.75.-k, 89.75.Da,  02.10.Ox}
\keywords{Kauffman model, random Boolean networks, relevant 
nodes, relevant components, gene regulation, number of attractors, stochastic process, combinatorics}
\maketitle

\section{Introduction}
\label{sec:intro}

Random Boolean networks are often used as generic models for the
dynamics of complex systems of interacting entities, such as social
and economic networks, neural networks, and gene or protein
interaction networks \cite{kauffman:random}. The simplest and most
widely studied of these models was introduced in 1969 by Kauffman
\cite{kauffman:metabolic} as a model for gene regulation.  The system
consists of $N$ nodes, each of which receives input from $K$ randomly
chosen other nodes. The network is updated synchronously, the state of
a node at time step $t$ being a Boolean function of the states of the
$K$ input nodes at the previous time step, $t-1$.  The Boolean
update functions are randomly assigned to every node in the network,
and together with the connectivity pattern they define the \emph{realization}
of the network. For any initial condition, the network eventually
settles on a periodic attractor.  Of special interest are
\emph{critical} networks, which lie at the boundary between a frozen
phase and a chaotic phase
\cite{derrida:random,derrida:phase,moreira:canalcritcond}.  In the
frozen phase, a perturbation at one node propagates during one time
step on an average to less than one node, and the attractor lengths
remain finite in the limit $N\to \infty$. In the chaotic phase, the
difference between two almost identical states increases exponentially
fast, because a perturbation propagates on an average to more than one
node during one time step \cite{aldana-gonzalez:boolean}.

During the last few years, great progress has been made in the
understanding of critical networks with $K=2$ inputs per node, see
\cite{bilke:stability,socolar:scaling,samuelsson:superpolynomial,gershenson:updating,klemm:asynchr,drossel:onnumber,kaufmanandco:scaling,ute:canal,samuelsson:damage}. %
They contain three classes of nodes, which behave differently on
attractors.  First, there are nodes that are frozen on the same value
on every attractor. Such nodes give a constant input to other nodes
and are otherwise irrelevant. They form the \emph{frozen core} of the
network. Second, there are nodes without outputs and nodes whose
outputs go only to irrelevant nodes. Though they may fluctuate, they
are also classified as irrelevant since they do not influence the
number and periods of attractors.  Third, there are \emph{relevant}
nodes, which belong to the nonfrozen nodes and lie on directed loops
built of links between relevant nodes, or they connect such loops by
chains of relevant nodes. These nodes determine completely the number
and period of attractors. A connected set of relevant nodes is called a
\emph{relevant component}. The relevant subnetwork consists of the
disjoint relevant components.  The nonfrozen nodes that are not
relevant sit on trees rooted in the relevant components.  Numerical
simulations and analytical estimation in \cite{socolar:scaling} showed
that the number of nonfrozen nodes scales in the limit $N\to\infty$ as
$N^{2/3}$ and the number of relevant nodes as $N^{1/3}$. An analytical
study of the distribution of attractor lengths in the limit of big
network size in \cite{samuelsson:superpolynomial} is shown in
\cite{drossel:onnumber} to contain the same result. The corresponding
probability distributions for the number of nonfrozen and relevant
nodes in terms of scaling functions have been obtained in
\cite{kaufmanandco:scaling,ute:canal} by means of combination of
analytical calculations and numerical simulations.  One application of
analytical approach developed in \cite{samuelsson:damage} extends the
findings of \cite{socolar:scaling} and gives slightly better numerical
approximations for the scaling functions.  In
\cite{kaufmanandco:scaling}, it was additionally shown that the number
of nonfrozen nodes with two relevant inputs scales as $N^{1/3}$, and
that the number of relevant nodes with two relevant inputs remains
finite in the limit $N\to\infty$. Probability distributions for these
two quantities were also given. From these results it follows that
just a few components are complex and include relevant nodes with two
relevant inputs, while most relevant components are simple loops whose
number is proportional to $\ln \sqrt{\bar N_{nf}}$ for large networks,
$\bar N_{nf}$ being the mean number of nonfrozen nodes.  The nonfrozen
part of a critical $K=2$ network is much like a critical network with
$K=1$ inputs per node and only the two nontrivial Boolean functions
``copy'' and ``invert''. All $N$ nodes in those networks are
nonfrozen, and the relevant nodes are arranged in simple loops. The mean
number of relevant nodes is proportional to $\sqrt{N}$
\cite{flyvbjerg:exact,samuelsson:damage}, the mean
number of loops equals $\ln \sqrt N$ for large networks, the number of
loops of length $l$ is Poisson distributed
with mean $1/l$ for $l\ll \sqrt{N}$\cite{drossel:onnumber}, the asymptotic probability
distribution for the number of relevant nodes is \cite{kaufmanandco:scaling}
$$
p_0(N_{rel}) 
= \frac{N_{rel}}{N}\; {\rm e}^{-N_{rel}^2/2N}\, .
$$

The number and period of attractors of a network is given by the
number and length of attractors of its relevant subnetwork. These are
obtained by combining the attractors of the individual relevant
components. If all relevant components were simple loops, it would
already follow that the number and length of
attractors increases with system size faster than any power law
\cite{drossel:number,drossel:onnumber}. 
Complex components make the number of attractors even larger, since
they possess themselves exponential numbers and periods of attractors
\cite{kaufman:on}. Therefore, complex components are important for the
understanding of dynamics of large random Boolean networks (RBNs).

Complex components are of interest also for other reasons. As shown in
\cite{fangting:yeast_rbn}, Boolean network models can be appropriate
models for the dynamics of genetic regulatory systems. Specialized
functional blocks should play an important role in such models, and
functional blocks can be modelled as complex components, when dynamics
can be treated as Boolean. Since the deterministic parallel update
rule used in many Boolean network models is not biologically realistic,
other, and in particular stochastic updating schemes have been
investigated
\cite{klemm:asynchr,greil:dynamics}. While they affect the number of attractors,
they do not influence the classification in frozen, nonfrozen, and
relevant nodes. The relevant components and the position of critical line
in the phase diagram \cite{aldana-gonzalez:boolean,moreira:canalcritcond} of Boolean networks are the
same for all update rules \cite{gershenson:updating}. Components and motifs (smaller units of nodes) 
play an important role at determining whether an attractor is robust
with respect to stochastic fluctuations \cite{klemm:robust_tplgy}.

Because of their importance for the dynamics of Boolean networks, we
study in this paper complex relevant components of RBNs with $K=2$.
 Future work will have to move
on to more realistic network structures.  In the following, we will
prove in section \ref{sec:theo} that in most large networks all nodes
with two relevant inputs reside in the same complex component, and we
will use this result to estimate analytically the number of complex
components in a network. Furthermore, simulation results are presented
and discussed. In section \ref{sec:p_of_m} we classify complex
components and discuss their probability distribution both
analytically and by using numerical simulations. Finally, in section
\ref{sec:motifs} we list all possible complex components with one and
two nodes with two relevant inputs and calculate their frequencies of
occurence. A summary of our findings is given in section
\ref{sec:final}.

%\section{Agglomeration}
\section{Number of relevant components}
\label{sec:theo}

In this section, we discuss the number of relevant components in
critical networks, with a special focus on complex relevant
components. We will first describe the stochastic procedure leading to
the relevant nodes and the components formed by them, which is the
basis of our computer simulations and analytical considerations.
Then, we will argue that most networks have not more than one complex
relevant component, and we will obtain the probability distribution
for the total number of relevant components.

\subsection{Algorithm for obtaining relevant nodes and relevant components}

In the subsequent sections, we will present numerical simulations of the ensemble
 of relevant subnetworks in a class of critical RBNs with two inputs
per node defined in the following way.
The probability $\beta$ to assign a frozen Boolean function (for which
the output value does not depend on the input values) to a node is equal
 to the probability to choose a reversible Boolean function (for which 
the output changes whenever any input value changes).
By $\gamma$ we denote the
probability of choosing a canalyzing function that yields one value for three
different input values combinations and once the other output value.
 The remaining proportion $1-2\beta-\gamma$ of functions are those that respond only
to one input.

We will use an efficient method presented in \cite{kaufmanandco:scaling},
which was developed for calculation of scaling properties of the number
of nonfrozen and of relevant nodes.
In \cite{kaufmanandco:scaling}, statistical properties of the ensemble of networks
were evaluated, and appropriate stochastic processes were used to determine the number of 
frozen and relevant nodes for a given network.
The frozen nodes were
determined by starting with nodes with frozen functions and by
iteratively determining nodes which become frozen due to being connected to
a frozen node. The outcome was the probability distribution
\begin{equation}
 G(y) \simeq  {\rm e}^{-y^3/2}(1-0.5\sqrt{y} + 3y)/(4\sqrt{y})\,
\label{gfit}
\end{equation}
for the variable 
 \begin{equation}
y = \frac{N_{nf}}{(N/\beta)^{2/3}}\, ,\label{defy}
\end{equation}
with $N_{nf}$ being the number of nonfrozen nodes. Most of these
nonfrozen nodes have only one relevant input, and the number of nonfrozen
 nodes with two relevant inputs $N_2$ is distributed according to
\begin{equation}
f(a) = \frac {2}{3a^{1/3}(1+\gamma/\beta)^{2/3}}G\left(\left(\frac a
{1+\gamma/\beta}\right)^{2/3}\right)\, , 
\label{fa}
\end{equation} with
$a = N_2/\sqrt{N_{nf}}$.  Then, only nonfrozen nodes and their
connections were considered. The idea for identification of relevant
nodes is to first identify nodes without outputs as irrelevant nodes
and then to determine iteratively all other nodes whose outputs go
only to irrelevant nodes. These nodes are also irrelevant.  The actual
stochastic procedure for obtaining relevant nodes in
\cite{kaufmanandco:scaling} is more involved, see original paper for
details.  For the scaling variable
\begin{equation}
\label{eq:z}
z=\frac{N_{rel}}{\left(\frac{N}{\beta+\gamma}\right)^{1/3}}\,
\end{equation} 
one gets a probability distribution $P(z)$ the shape of which depends on
the parameters $\beta$ and $\gamma$. $P(z)$ decays exponentially
for large $z$. An explicit expression for $P(z)$ in terms of (\ref{fa})
and another scaling function occuring in the stochastic process
 is given in Eq.~(29) of
\cite{kaufmanandco:scaling}.

 Most relevant nodes have one relevant
input. The full ensemble probability $\tilde P(m;z)$ for (\ref{eq:z})
 to be in the range $[z,z+dz]$, while $m$ of the relevant nodes
 have two relevant inputs,  
depends again on $\beta$ and $\gamma$.
An important observation is that $\tilde P(m;z)$ does not depend on $N$
in the limit of large network size. The mean number of
relevant nodes with two relevant inputs is therefore finite in this limit.

The described procedure from \cite{kaufmanandco:scaling} provides us with
 the number of relevant nodes $N_{rel}$
 and  the number of relevant nodes with two relevant inputs $m$.
Connecting these nodes randomly generates a network from the ensemble of relevant
subnetworks. A  finite fraction $\epsilon$ of
relevant nodes are arranged in simple loops. Simple loops of length $ l$
appear \cite{kaufmanandco:scaling} with probability
\begin {equation}
P(l)=
1/l \text{\ for\ } l<l_c \text{\ with\ } l_c\sim N_{rel}\, .
\label{eq:pofl}
\end{equation}
 in the limit $N_{rel} \to \infty$.
 The other $(1-\epsilon)N_{rel}$ relevant nodes sit on complex
components containing nodes with two relevant inputs.

\subsection{Number of complex relevant components}

We argue in the following that most relevant nodes with two relevant inputs
reside in one component. We calculate the probability that this is not
the case and show it to be small and to decrease as $1/m$ for large
$m$.

We neglect the cases where one relevant node has two relevant inputs and two
relevant outputs, as well as where a node
 has more than two relevant outputs.
The probability for such nodes vanishes in the limit of large network size.
Then, the number of relevant nodes with two relevant
outputs is identical to the number of relevant nodes with
two relevant inputs $m$.

In the following, we first fix the value of $m$. Later, we will perform
summations over $m$ using the  probability distribution $p(m)$.
For a given $m$, the ensemble of relevant networks can
be constructed by connecting the $2m$ nodes with 2 relevant inputs or
two relevant outputs amongst one another, each realization of
connections appearing with equal probability, and by subsequently
inserting the nodes with one relevant input into the existing connections. The
probability that the $m$ nodes are not in the same component is thus
the probability that the $2m$ nodes do not form a fully connected
directed graph. We denote this probability by $G_m$. The strategy for
calculating $G_m$ is the following. We look at all possible
partitionings of connected subgraphs of the $2m$ nodes with
connections and sum the contributions to the probability coming from
$1,2,3, \ldots$ fully connected subgraphs. Each of those contributions
itself includes a sum of different partitionings with a fixed number of
subgraphs. The last contribution in the overall sum comes from a
partitioning in $m$ connected pairs of nodes, one with two inputs and
one output and one with two outputs and one input. If we do this, we
get $G_1=0$, $G_2=0.1$, $G_3=0.1$, $G_4\approx 0.08$, $G_5\approx
0.07$. The general formula reads
%\begin{equation}
\begin{align}
&N_m = (m!)^2\sum_{\text{partitions\ }\{n_i|i\}} \prod_i \frac{((3i)!-N_i)^{n_i}}{n_i!(i!)^{2n_i}}\,,\\\nonumber
&\hbox{ with } N_m = (3m)!\;G_m\,,
\end{align}
where the sum is over all nontrivial partitions of $m$ nodes in $n_i$ groups of $i$ nodes, $m=\sum_i in_i$.
To estimate the asymptotic behavior for large $m$, we note that the largest contribution for large $m$ comes from the partitioning in one connected pair and a fully connected rest. We approximate $G_m$ by the contribution from a partitioning in one connected pair and an arbitrarily connected rest: $6m^2 (3(m-1))! / (3m)!\to 2/(9m)$ for large $m$. We conclude that all nodes with two relevant inputs are usually gathered in one complex component, even for small values of $m$.

\subsection{Total number of relevant components}

We build the relevant components by first ignoring one input of each
nonfrozen node that has two inputs. The second inputs will be connected
later. We thus first build a $K=1$ critical network from the nonfrozen
nodes, many properties of which are known from the literature.  In
particular, all nonfrozen nodes will at this stage be arranged in
\begin{equation}
\label{eq:C}
C \simeq 1/2(\ln(2N_{nf})+\gamma_E)\,,
\end{equation}
simple loops and trees rooted in loops \cite{samuelsson:maps}.
$\gamma_E\approx 0.577$ is the Euler-Mascheroni constant.  The width
of the distribution of the number of loops is asymptotically
\begin{equation}
\label{eq:sigma}
\sigma^2=C-\pi^2/8\, .
\end{equation}
\comment{\sqrt{pi/2}sqrtNnf is asympt exact number of nodes on circles und is proportional to biggest loop length! => \sqrt{2/pi}sqrtNnf is proportional
to tree size!}

Equations (\ref{eq:C}), (\ref{eq:sigma}) are valid for fixed values of
$N_{nf}$. If we insert for $N_{nf}$ the mean value $\bar N_{nf}\approx
0.62 (N/\beta)^{2/3}$ (see equation (12) in
\cite{kaufmanandco:scaling}),  we get $\bar C\approx 6.4$ for $N=2^{23}$. 
%Umständlicherer Weg:
% To determine a better approximation for $\bar C$ from the combined
%distribution for both $N_{nf}$ and $C$ numerically is an elaborate
%task, because the PD for the number of simple loops (\ref{eq:C}),
%which is found in \cite{samuelsson:maps} in (II.C15), includes
%sign-less Stirling coefficients in a complex expression
%$P^\prime_{N_{nf}}(C=\mu)=\sum_{n=\mu}^{N_{nf}}\binom{N_{nf}}{n}\left[\begin{smallmatrix}n\\
%\mu\end{smallmatrix}\right]n/{N_{nf}^{n+1}}$.  To estimate the
%corrections due to usage of combined shark-fin-PD, we
If we use the
full probability distribution for the number of nonfrozen nodes and
the number of simple loops, we obtain the following formula for the
mean number of simple loops:
\begin{eqnarray}
\label{eq:Cfull}
\bar C(N)&\simeq& \sum_{\mu=1}^N \mu \sum_{L=1}^N \frac{1}{L!}\begin{bmatrix}L\\ \mu\end{bmatrix}
\sum_{N_{nf}=1}^N (N/\beta)^{-2/3}\nonumber \\ &&  G(N_{nf}(N/\beta)^{-2/3}) \binom{N_{nf}}{L}\frac{L L!}{N_{nf}^{L+1}}\;,
\end{eqnarray}
which is a combination of exact results (II.C16) and (II.C17) for the
ensemble of $K=1$ networks from
\cite{samuelsson:maps}, and of (\ref{gfit}). We designed a program to
effectively evaluate (\ref{eq:Cfull}) numerically. For the
above-mentioned system size $N = 2^{23}$  we get $\bar
C$=6.13. We use this value in the following discussion. Note that
using mean values instead of full probability distributions gives good
estimations of the order of magnitude in the present context.

Now we estimate the number of components out of the $\bar C$ simple
loop components on which the relevant nodes with two relevant inputs
and the nodes with two relevant outputs sit. By assuming that all of
them will end up on the same complex component when the second inputs
will be connected, we obtain our desired result for the mean number of
relevant components. In the following, we approximate
the mean value of some functions by the
functions of the mean values, so that the results (\ref{eq:nk}) and
(\ref{eq:Crel}) are valid approximately.

Because of (\ref{eq:pofl}), the mean number of components with loops
of lengths from logarithmically constant intervals will be
constant. If $l_1$ and $l_2$ denote the lower and upper boundary of an
interval, there is on an average one loop in an interval of logarithmic
 size $l_2/l_1 = {\rm e}$.
We consider such intervals with the upper boundaries $l_2$ equal to ${\rm e}$, ${\rm e}^2$, and so on.
For two neighboring intervals, the factor between the numbers of nodes on the loops
is ${\rm e}$ on an average.
Since for an interval the total number of nodes on the loops and on their trees
is proportional to the number of nodes on the loops, the factor between the total numbers of nodes in
two neighboring intervals is also ${\rm e}$.
With this assumption, the proportion of nodes in the $p$th largest component is
$\lambda {\rm e}^{-(p-1)}$, with $\lambda = 1-1/{\rm e}\approx 0.632$. We assumed that ${\rm e}^{-\bar C}\ll 1$.
The value $\lambda = 1-1/{\rm e}$ is not far from the numerically determined value $\lambda\approx 0.624$ for
the proportion of nodes in the largest component in ensembles of $K=1$ networks with any large but fixed
number of nonfrozen nodes.
Let us now turn to the $m$ relevant nodes with two relevant inputs, with one input being ignored.
The number of nodes with two relevant inputs or two relevant outputs
in the largest nonfrozen component is then of the order $2m \lambda$,
and the smaller components will contain $2m\lambda /{\rm e}$, $2m\lambda
/{\rm e}^2$,\ \dots,\ $1$ nodes with two relevant inputs. For the
purpose of estimation, we neglect further the fact that the distribution of
the number of nodes with two relevant inputs $p(m)$ depends
\cite{kaufmanandco:scaling} on the number of relevant nodes and,
therefore, correlates with the number of nonfrozen nodes on the loops.

The total number $n(m)$ of nonfrozen components with nodes with two
relevant inputs (one of them is cut off) is consequently obtained from
the condition $2m\lambda\;\!{\rm e}^{-(n(m)-1)}=1$ for $m\ge 1$. For $m=1$,
this gives $n(1) \simeq 1.234$. The exact result $n(1) = 4/3$ is
within 10 percent of this estimate.
To explain the exact result for fixed $m=1$ we consider all possible relevant
components containing the one node with two relevant inputs, which result after
connecting the cut off input. They are represented in figure~\ref{fig:m1motifs}.
Clearly, the two relevant inputs could have been cut off with the same probability.
As explained in section \ref{sec:motifs}, relative frequency
of occurence for the second component in figure~\ref{fig:m1motifs} is $1/3$, and
cutting one input off the node with two inputs leads to one nonfrozen component.
Such components have therefore resulted from one nonfrozen component. For the
first component in figure~\ref{fig:m1motifs}, which appears with relative
probability $2/3$, cutting one input leads to two nonfrozen components with
probability $1/2$. The mean number of $K=1$ nonfrozen components connected by the relevant
node with two relevant inputs is then $1*2/3+2*1/3=4/3$.
\comment{naive distribution of one node with 2 inputs and one with 2 outputs over
nf-components with the above size distribution is wrong since f.i.
<\lambda*(\lambda/e)>!=<\lambda><\lambda/e>
}

\begin{figure}
\begin{center}
\vskip 0.5cm
\includegraphics*[width=0.95\columnwidth]{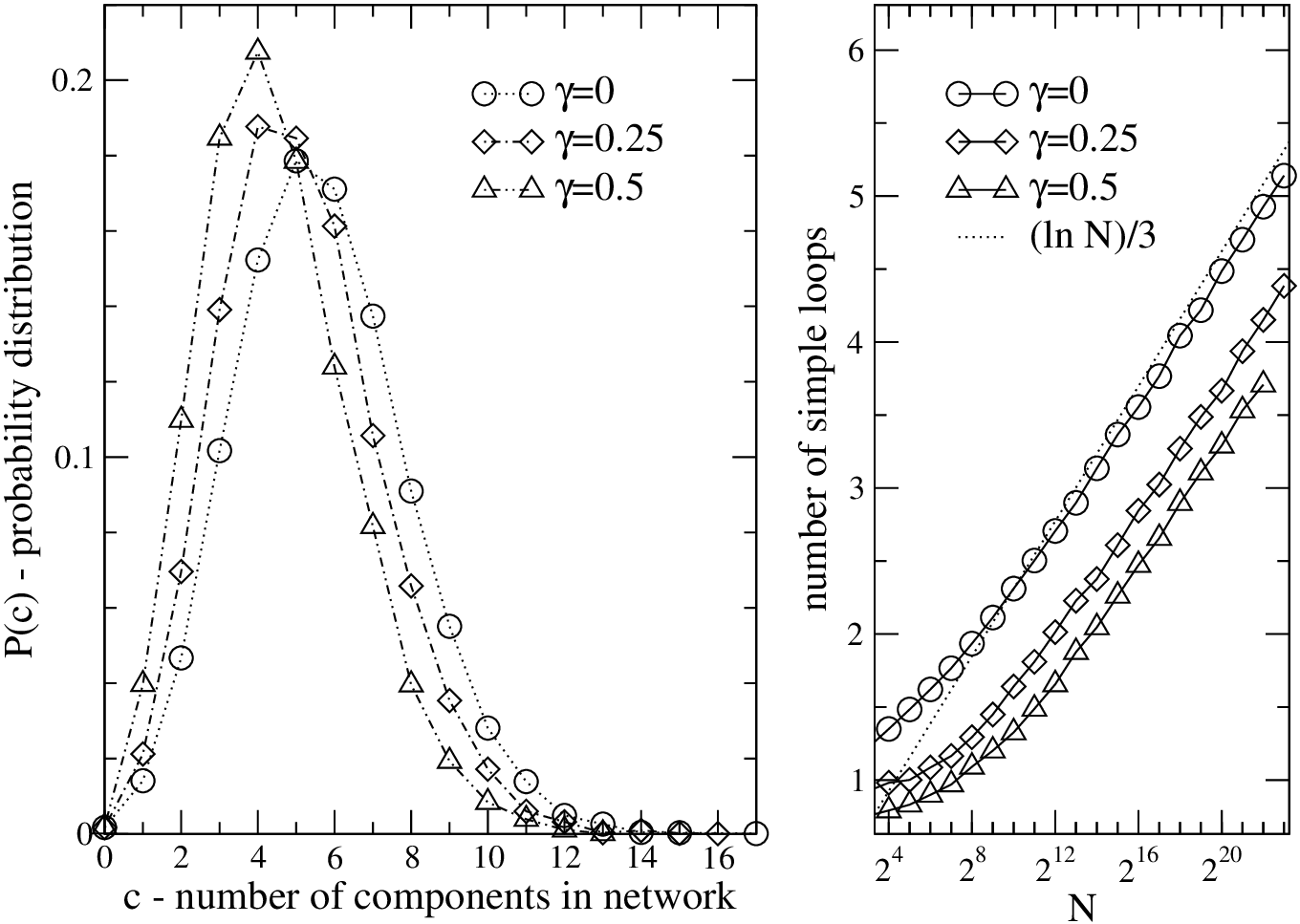}
\end{center}
\caption{\emph{Left figure:} Probability distribution for the number of relevant components per network. Results were averaged over $10^4$\comment{8000?} network realizations for the following model parameters: $\beta=1/8$; $\gamma=0,1/4,1/2$; $N=2^{23}\approx 8\cdot 10^6$ for $\gamma\ne 1/2$ and $N=2^{22}$ for $\gamma=1/2$. Obtained in simulations results for mean numbers of components for $\gamma=0,1/4,1/2$ are $5.7$, $5.1$, $4.5$ respectively. Using results of simulations or analytical results for $p(m)$, see section \ref{sec:p_of_m} for details, and (\ref{eq:Crel}) we get $5.5$, $4.5$, $3.6$. The agreement between theoretical estimation and simulations is satisfactory. %
Half maximal widths of the simulation curves are $5.3$, $5.0$, $4.6$ compared to values $4.43$, $4.43$, $4.32$, determined from (\ref{eq:sigma}), whereby $\bar m$-dependence has been ignored.
\comment{1e4 nets,
For gamma=0($Nini=8e6$),0.25(8e6),0.5(4e6) we have respectively
x-barycenter i.e. <c>: 5.67,5.1,4.48; half maximal width 5.27,4.94,4.56;
c of maximum P(c): 5,4,4;
}%
\\\emph{Right figure:} Number of simple loop components per network in $K=2$ critical RBNs for the same sets of model parameters and numerical simulation parameters as in the left figure. Our simulation results for the mean number of simple loop components in the largest considered networks for $\gamma=0,1/4,1/2$ are $5.1$, $4.4$, $3.7$ respectively. From the estimation (\ref{eq:Crel}) we get
\comment{the above estimation values minus 1 is wrong, since with probability 0.51,0.32,0.27 there are no relevant nodes with two relevant inputs and nothing has to be subtracted
}%
$5.0$, $4.2$, $3.3$, since all nodes with two relevant inputs are assumed to reside in one complex component as explained in the main text.
}
\label{fig:compscnt}
\end{figure}

Taking all results together, we obtain the following estimate for the
mean number of nonfrozen components with relevant nodes with two relevant
inputs: 
\begin{equation}
\label{eq:nk}
\bar n = \sum_{m=1}^N p(m) n(m)\approx
\sum_{m=1}^N{ p(m)\ln{(2m({\rm e}-1))} }\;.
\end{equation}
We recall that (almost) all nodes with two relevant inputs are gathered in one complex component, and
 with (\ref{eq:nk}) we get an estimation for the mean number of relevant components, which is valid for large networks:
\begin{multline}
\label{eq:Crel}
C_{rel}^{K=2}(N) \approx \bar C(N)  - \sum_{m=1}^N p(m)(n(m)-1)=\\
 \bar C(N) - \;\overline{\ln m}\; - (1-p(0))\ln(2\lambda) \,.
\end{multline}
We support (\ref{eq:Crel}) by evaluating the number of relevant components in numerical simulations, see \hbox{figure~\ref{fig:compscnt}}. For comparison with (\ref{eq:Crel}), some results from section \ref{sec:p_of_m} have been used, see \hbox{figure~\ref{fig:pvonm}} there.

The average number of complex
components per network can be extracted from results presented in
\hbox{figure~\ref{fig:compscnt}} as the difference between the number of
relevant components and the number of simple loop components. For
model parameters $\beta=1/8$ and $\gamma=0,1/4,1/2$ we get $0.6$,
$0.7$ and $0.8$ complex components on average respectively. For the
three cases, in section \ref{sec:p_of_m} we show that the probability
to have no nodes with two relevant inputs in a network is $0.51$,
$0.32$ and $0.27$, so that the average number of complex components in
the networks with complex components evaluates to $1.2$, $1.0$ and
$1.1$ respectively, justifying the assumption that there is usually no
more than one complex component. Statistical errors of simulation
results consistently allow for errors in the first digit after the
comma.

The difference between simulation results in figure~\ref{fig:compscnt} and
the results from (\ref{eq:Crel}) is smaller for smaller values of $\gamma$,
where the approximations used to obtain (\ref{eq:Crel}) are intuitively
better. An estimation in the spirit of (\ref{eq:nk}) and (\ref{eq:Crel}), which would
take into account that the probability for relevant nodes with two relevant
inputs to be arranged in more than one relevant complex component is small but
greater than zero, would not change (\ref{eq:nk}), but it would lead to slightly
smaller values of $C_{rel}^{K=2}(N,\beta,\gamma)$ in (\ref{eq:Crel}).
This observation partially explains the difference between the simulation and the
estimation results in figure~\ref{fig:compscnt}.

%useful general comments
\comment{the probability for a nf node to be relevant is 1/sqrtNnf. Beweis: betrachte größte comp mit sqrtNnf=sqrtM loop-nodes und trees der groesse sqrtM. addiere ein knoten mit 2 inputs: er ist meist auf einem baum. moeglichkeiten, dass er relevant ist (``nach oben angeschlossen wird'') sind sqrtM*(sqrtM+(sqrtM-1)+(sqrtM-2)+...+1)=M*sqrtM/2. Insgesamt moeglichkeiten anzuschliessen: M*M->WSK=1/2sqrtM. BTW wie in einem comment oben schon erwähnt, average tree size is $\sqrt{N_{nf}}$
}
\comment{for nonfrozen components with Nnf nodes, ~sqrtNnf on the loops, tree size each ~sqrtNnf, number of nodes with 2 inputs ~sqrtNnf, probability for them to be relevant is ~\bar m/sqrtNnf. Somehow I thought that (not checked again)
the m's sit probably on nonfrozen trees in the distance ~ln Nnf from the root, if we imagine the tree as each node having two outputs.
}
\comment{
It is not enough to say that relevant nodes are non-constant and control at least one other relevant node, since for a single loop we need to extra define one node as relevant. Also for two loops with a link between them.
}

\section{Distribution of complex components}
\label{sec:p_of_m}

The findings of the last section together with the results in
\cite{kaufmanandco:scaling} enable us to evaluate the mean number of
relevant nodes with two relevant inputs,
the probability distribution for the number of relevant nodes with two
relevant inputs, and the mean number of relevant complex components per
network with a given number of nodes with two relevant inputs.
The complex components have nontrivial dynamics and can be associated with functional blocks in
real genetic networks.

Let us first evaluate the number of relevant nodes with two
inputs. Using Equations (23), (24), (29), (31) from \cite{kaufmanandco:scaling},
we can write its probability distribution as
\begin{eqnarray}
p(m) &=& \int_0^\infty \tilde P(m;z) dz \nonumber\\
&=&  \frac{1}{(m!)^2}\int_0^\infty dz \frac{(z^3/2)^m  P(z)}{I_0(\sqrt{2z^3})}\,,  \label{eq:pofm}
\end{eqnarray}
where $I_n(z)$ denotes the modified Bessel function of the first kind
and satisfies the relations $I_0(x) = \sum_{m=0}^\infty
(\frac{x^m}{2^m m!})^2$ and $I_1(x) = 2/x\sum_{m=0}^\infty
m(\frac{x^m}{2^m m!})^2$.

\begin{figure}
\begin{center}
\vskip 0.5cm
\includegraphics*[width=0.95\columnwidth]{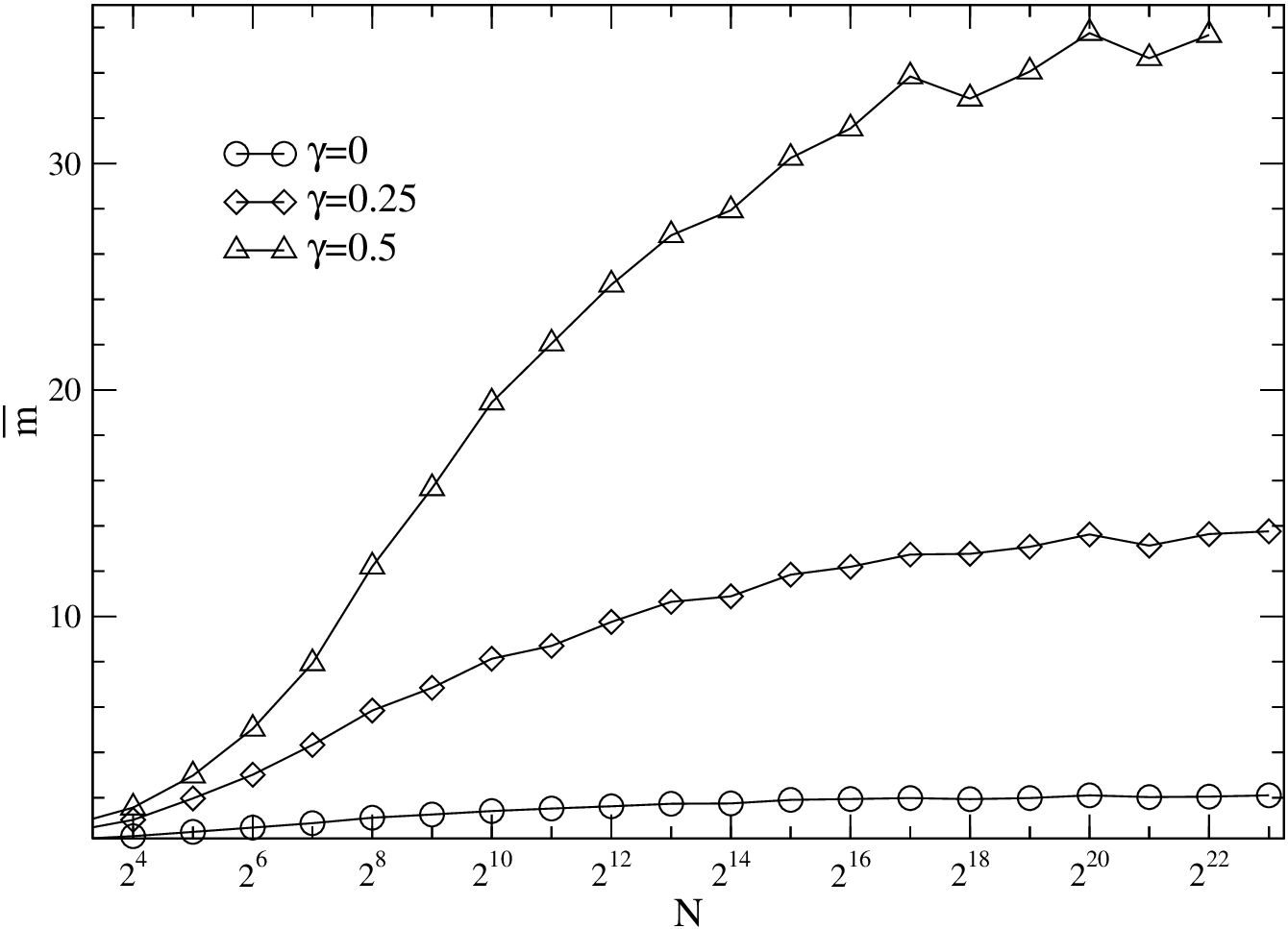}
\end{center}
\caption{The mean number of nodes with two relevant inputs for $\beta=1/8$ as function of the network size $N$. The averaging was performed over $10^4$\comment{8000?} network realizations.}
\label{fig:twoinp}
\end{figure}

The mean number of nodes with two relevant inputs is then finite and is given by
\begin{equation}
\bar{m} = \frac{1}{2}\int_0^\infty dz \frac{\sqrt{2z^3}\; I_1(\sqrt{2z^3})}{I_0(\sqrt{2z^3})} P(z)dz\,.
\end{equation}
The integral converges since $ P(z)$ decreases exponentially  for large $z$ (compare \cite{kaufmanandco:scaling}), and the rest of the integrand grows faster than linearly but slower than quadratically with $z$.  figure~\ref{fig:twoinp} shows results of computer simulations for the values of 
$\bar{m}$. With increasing $N$, they approach a constant value.

For further convenience, we introduce the notation
$\kappa=1+\gamma/\beta$.  The dependence of $\bar{m}$ and $p(m)$ on
the model parameters $\beta$ and $\gamma$ is fully determined by the
dependence of $P(z)$ on $\kappa$. Since $P(z)$ becomes broader for
larger $\kappa$ (therefore assuming smaller values at small $z$), see
\cite{kaufmanandco:scaling}, $\bar m$ increases with $\kappa$. This
means that increasing $\kappa$ leads
to more relevant nodes with two relevant inputs. We will look at the
dependence of $p(m)$ on $\kappa$ in more detail later.
\comment{$P(z)$ depends on $\kappa$ through the dependence of $f(a)da=G(y)dy$
on $\kappa$. After variable substitution $a=\kappa y^(3/2)$ the function
$P(z)$ depends on model parameters only through $\kappa$ (which sits then
in the $C_a(z/a^0.333)/a^0.333$) and
$\bar a = \kappa * const$
By the way, $\kappa$ lies in the range from $1$ to $\beta N_{ini}^0.5$
($\approx\infty$).
}

\begin{figure}
\begin{center}
\vskip 0.5cm
\includegraphics*[width=0.95\columnwidth]{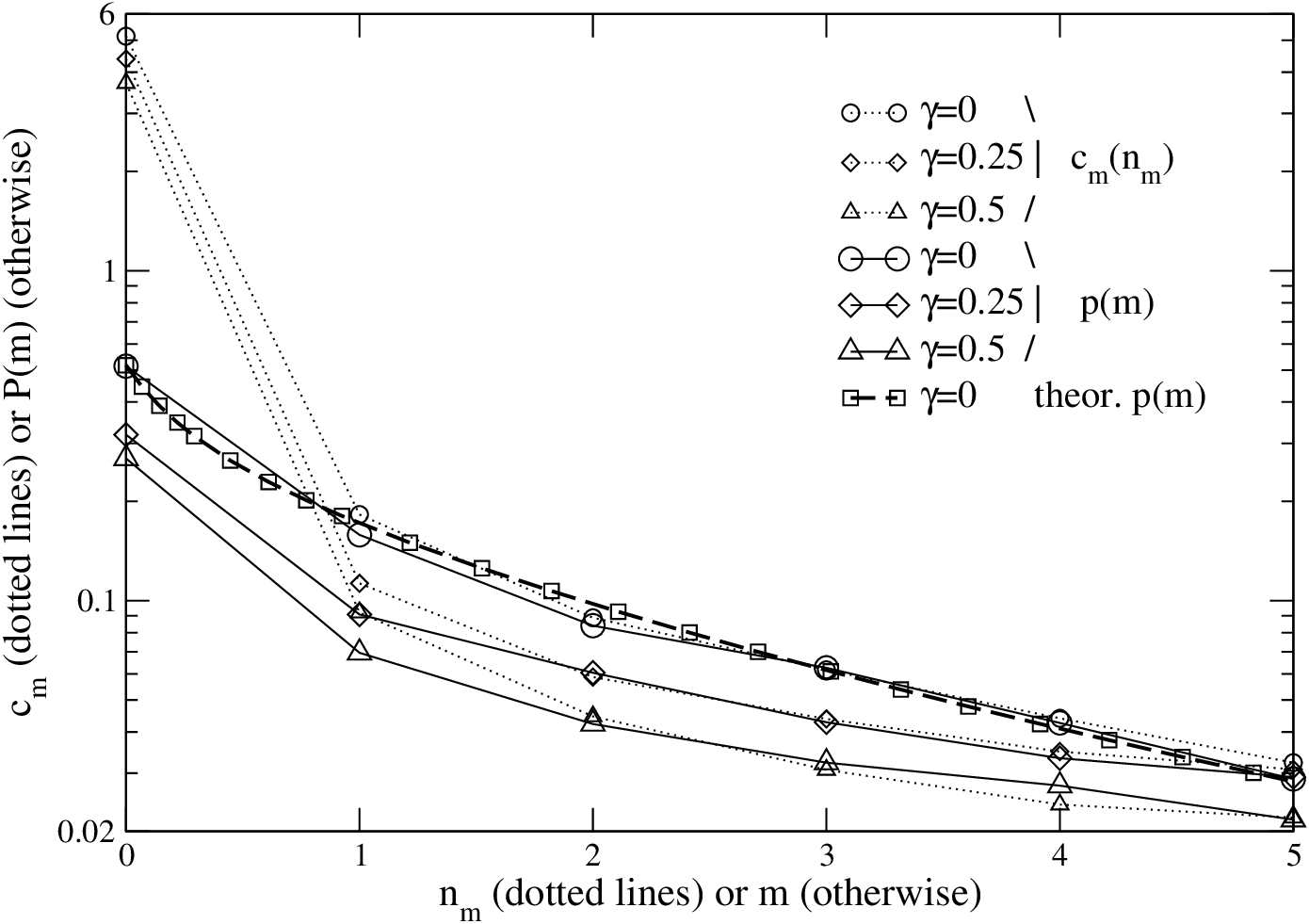}
\end{center}
\caption{Simulation results for the number $c_m(n_m)$ of components per network with $n_m$ nodes with two relevant inputs (dotted lines). Simulation results for the probability distribution $p(m)$ for the number $m$ of nodes with two relevant inputs per network (solid lines). Theoretical prediction for the probability distribution $p(m)$ (dashed line). The model parameters used are $\beta=1/8$ and $\gamma=0$ (circles, squares) or $\gamma=1/4$ (rhombi) or $\gamma=1/2$ (triangles). All simulations were run with 8000 different network realizations and results were averaged afterwards. Network size was $N\sim 10^7$ (dotted lines), $N\sim 10^5$ (solid lines). Approximation $P^{fit}_{\gamma=0}(z)=0.62\;{\rm e}^{-0.65z}$ was used to obtain the dashed curve.
}
\label{fig:pvonm}
\end{figure}
\begin{figure}
\begin{center}
\vskip 0.5cm
\includegraphics*[width=0.95\columnwidth]{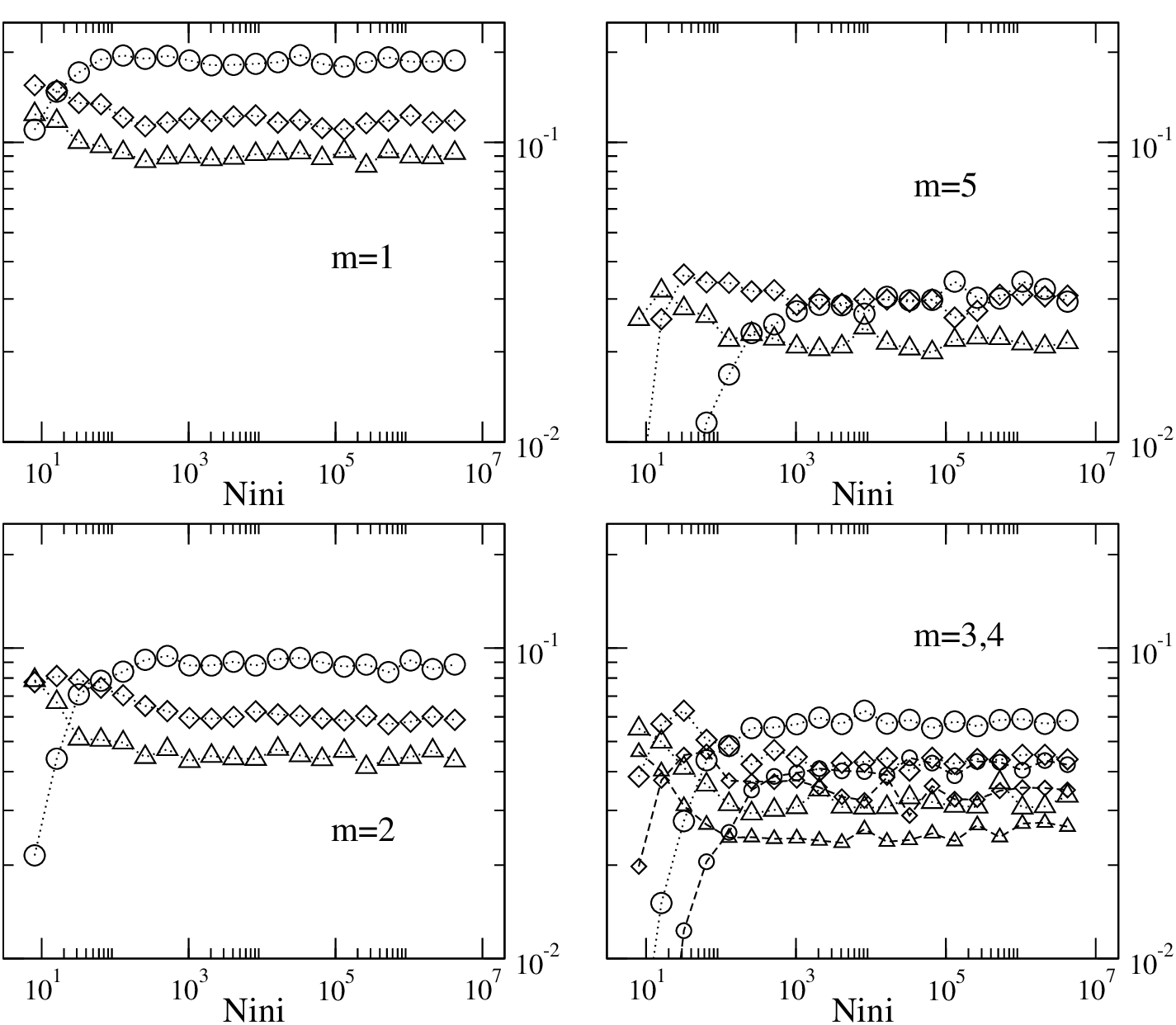}
\end{center}
\caption{
The number of relevant complex components with  $m$  nodes with two relevant inputs is constant for large networks. In the figure, circles, diamonds and triangles correspond to model parameters $\beta=1/8$ and $\gamma=0, 1/4, 1/2$ respectively. All simulations were run with 8000 different network realizations.
}
\label{fig:pvonm_convergence}
\end{figure}

We turn to the discussion of the connection between $p(m)$ and the
number of components with a given number of nodes with two relevant
inputs. From section \ref{sec:theo} we know that, except for a small
correction, most nodes with two relevant inputs reside in one complex
component. Therefore, the fraction of number of networks, where we
find a given value of $m$ will simultaneously be the fraction of
number of networks where we find complex components (one complex
component) with this number of nodes with two relevant inputs. In
figure~\ref{fig:pvonm}, the dotted lines represent the mean number $c_m$
of components per network with a given number $n_m$ of nodes with two
relevant inputs, obtained in our computer simulations by counting the
number of such components in the generated network ensemble for
different model parameters.  To verify that the relevant nodes with
two relevant inputs usually sit in the same component, we also
evaluated the probability distribution $p(m)$ numerically (solid lines
in figure~\ref{fig:pvonm}).  For $m>1$ and $n_m>1$, the difference
between the solid and the corresponding dotted lines in
figure~\ref{fig:pvonm} is mainly due to statistical fluctuations. They
become larger for larger networks, we therefore used comparatively
small networks with $N\sim 10^5$ to obtain the solid lines.  In the
rare network realizations with two complex relevant components
(appearing more often with increasing $\bar m$) the smaller complex
component has mostly one relevant node with two relevant inputs
($m=1$), see the data points for $m=n_m=1$ in figure~\ref{fig:pvonm}.
Finally, we calculated $p(m)$ analytically with (\ref{eq:pofm}).  The
theoretical dashed curve in figure~\ref{fig:pvonm} agrees well with the
simulation results. We only calculated analytical results for $\gamma=0$.
That the number of components with a
given value of $m$ is indeed independent of the system size is shown
in figure~\ref{fig:pvonm_convergence}.

To complete our understanding of simulation results in figure~\ref{fig:pvonm} we point out the connection to the results from section \ref{sec:theo}. For $m=0$ the solid lines represent the fraction of networks without nodes with two relevant inputs. We used the results for these fractions $0.51$, $0.32$, $0.27$ for figure~\ref{fig:compscnt}. Network realizations without relevant nodes with two relevant inputs are indistinguishable from networks in the model with $K=1$. In this context we want to mention that the fraction of ``frozen'' networks without relevant nodes $\sim(\beta/N)^{1/3}$
\comment{for N_rel=0..1 we have z=0..(beta*kappa/N)**(1/3). Integral[dz const/kappa**0.333] in 0..(beta*kappa/N)**(1/3) gives ~N**(-1/3) 
}%
is negligible in the limit $N\to\infty$, whereas $\lim_{z\to 0}
P(z)\ne 0$.  The dotted curves at $n_m=0$ represent the number of simple loops per
network. For $\gamma=0$, $0.25$, $0.5$ this number is $5.1$,
$4.4$, $3.7$ respectively. These values are in good agreement with
those calculated with (\ref{eq:Crel}), see figure~\ref{fig:compscnt}.
\comment{
%and with mean values $\bar m$ taken from figure~\ref{fig:twoinp}: $4.89$, $4.36$, $3.8$.
I have not considered \bar(\ln m) as \ln(\bar m), the values which stood at this place were obsolete rests from
a previous version of the article, sorry about this :)
}%

Let us conclude this section by discussing the dependence of $p(m)$ on
the parameter $\kappa$. For fixed $\kappa$, $p(m)$ is a monotonously
decreasing function, which is flatter for larger $\kappa$. For fixed
$m$, the value $p(m)$ as function of $\kappa$ has a maximum, the
position of which moves to larger $\kappa$ with increasing $m$. We see
an indication of this in figure~\ref{fig:pvonm} and
figure~\ref{fig:pvonm_convergence}, where the data points for $\kappa=1$
are not the highest ones when $m$ is larger than 4.  Stronger support
comes from an analysis of Eq.~(\ref{eq:pofm}).  The integrand apart
from $P(z)$ has one global maximum, which shifts to larger values of
$z$ as $m$ increases. The function $P(z)$ is monotonically decreasing,
being flatter and starting at a smaller value for larger $\kappa$.
From these properties of the two contributions to the integrand, we
find immediately that the value $p(m)$ at a given $m$ as function of $\kappa$
has a maximum, which moves to larger $\kappa$ with increasing $m$.
For very large $\kappa$, the function $P(z)$ in (\ref{eq:pofm})can be
approximated \cite{kaufmanandco:scaling} by its value
$P(z=0)=\sqrt{2\pi}/(4\kappa^{1/3})$ for not too large $m$, revealing
how $p(m)$ decreases with increasing $\kappa$.

From these considerations it follows that complex components with a
small number of nodes with two relevant inputs will appear with
smaller probability if we increase $\kappa$, while complex components
with larger $m\sim\bar m$ occur more often.

\section{Relative frequencies of topologically different components}
\label{sec:motifs}

So far, we characterized relevant complex components by the number of nodes with two relevant inputs found in them. We determined, which complex components and with what probability are found in a network asymptotically. Now, we want to consider the topology of such complex components.

We consider two complex components as different if they possess a
different topology. Components with the same topology but with
different numbers of nodes in the different chains of relevant nodes
with one relevant input, have similar dynamics, their main difference
being different time delays along chains of nodes. From our study of
simple examples of complex components in \cite{kaufman:on}, we know
that changing the length of these chains does not change the types of
attractors and the asymptotic dependence of the mean number and length
of these attractors on the component size.  Therefore, we restrict
ourselves to considering all possible different interconnections of
relevant nodes with two relevant inputs with relevant nodes with
two relevant outputs, i.e., the different
topologies, irrespective of the lengths of the chains between them.

\begin{figure}
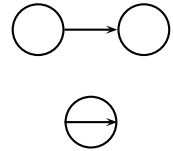

\begin{center}
\vskip 0.5cm

\psset{unit=1pt,fillcolor=black}
%\rput(-80,-25){\makebox[0pt][c]{$32 \times$}} %B
\pscircle(-30,-25){10}
\pscircle(10,-25){10}
\psline{->}(-20,-25)(0,-25)
\pscircle(-10,-60){10}
\psline{->}(-20,-60)(0,-60)
\vspace*{60pt}
\end{center}
\caption{Schematic representation of the two topologically different relevant complex components with two nodes with two relevant inputs. Arcs stand for one link or a chain of nodes with one relevant input. Each crossing of arcs marks a node with two relevant inputs or outputs. Arrows depict the inputs of nodes with two relevant inputs, they appear only where they are important to distinguish the complex component. The first component appears in a network twice as often as the second one.}
\label{fig:m1motifs}
\end{figure}

We first consider complex components with one node with two relevant
inputs, that is the case $m=1$. The two different possibilities to
construct these components are the ones studied in
\cite{kaufman:on} and schematically shown in
figure~\ref{fig:m1motifs}. In our simulations, the relative probability
of obtaining two simple loops with a chain of nodes between them in a
relevant network is $2/3$, and the relative probability of obtaining a
simple loop with an extra link is $1/3$. For small
values of $m$, these relative frequencies of the different possible
complex components can easily be determined exactly. One has simply to count
the number of different ways to connect the nodes with two relevant
inputs or outputs out of the total number $(3m)!$ leading to the
considered component. For the case of figure~\ref{fig:m1motifs}, the
number of ways to construct the two components are $4$ and $2$
respectively.

\begin{figure}
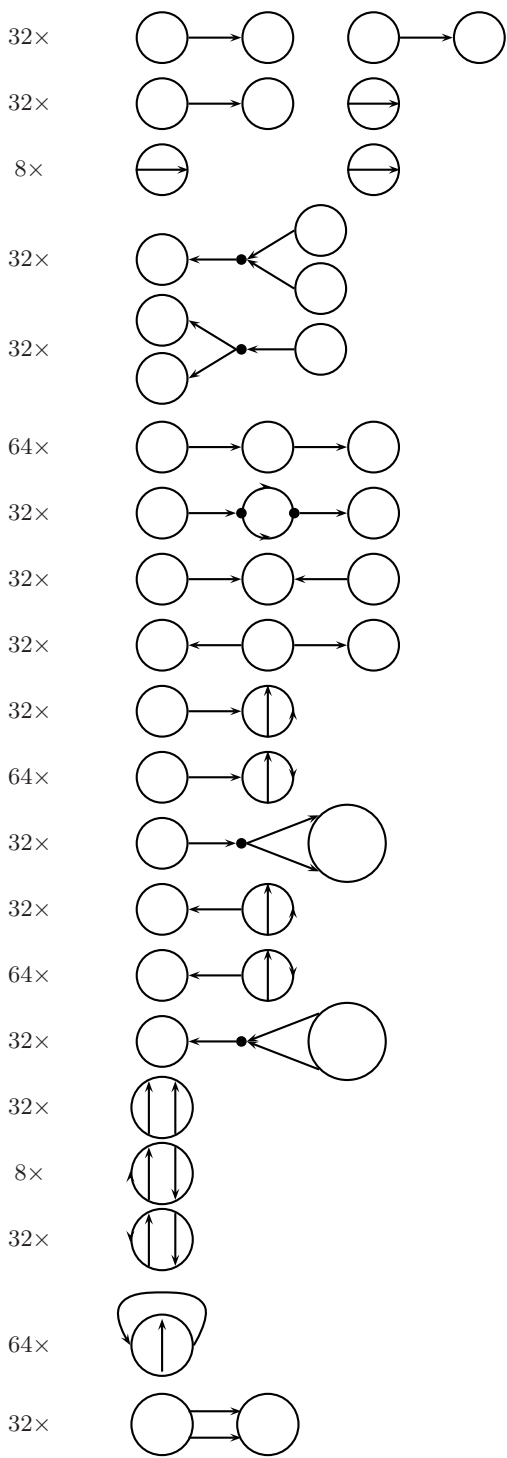

\begin{center}
\vskip 0.5cm

\psset{unit=1pt,fillcolor=black}

\rput(-80,0){\makebox[0pt][c]{$32 \times$}} %A
\pscircle(-30,0){10}
\pscircle(10,0){10}
\psline{->}(-20,0)(0,0)
\pscircle(50,0){10}
\pscircle(90,0){10}
\psline{->}(60,0)(80,0)
\rput(-80,-25){\makebox[0pt][c]{$32 \times$}} %B
\pscircle(-30,-25){10}
\pscircle(10,-25){10}
\psline{->}(-20,-25)(0,-25)
\pscircle(50,-25){10}
\psline{->}(40,-25)(60,-25)
\rput(-80,-50){\makebox[0pt][c]{$8 \times$}} %C
\pscircle(-30,-50){10}
\psline{->}(-40,-50)(-20,-50)
\pscircle(50,-50){10}
\psline{->}(40,-50)(60,-50)
\rput(-80,-84){\makebox[0pt][c]{$32 \times$}} %D
\pscircle(-30,-84){10}
\psline{<-}(-20,-84)(-2,-84)
\pscircle[fillstyle=solid](0,-84){2}
\psline{->}(20,-73)(2,-84)
\psline{->}(20,-95)(2,-84)
\pscircle(30,-73){10}
\pscircle(30,-95){10}
\rput(-80,-118){\makebox[0pt][c]{$32 \times$}} %E
\pscircle(30,-118){10}
\psline{->}(20,-118)(2,-118)
\pscircle[fillstyle=solid](0,-118){2}
\psline{<-}(-20,-107)(-2,-118)
\psline{<-}(-20,-129)(-2,-118)
\pscircle(-30,-107){10}
\pscircle(-30,-129){10}
\rput(-80,-155){\makebox[0pt][c]{$64 \times$}} %F
\pscircle(-30,-155){10}
\pscircle(10,-155){10}
\pscircle(50,-155){10}
\psline{->}(-20,-155)(0,-155)
\psline{->}(20,-155)(40,-155)
\rput(-80,-180){\makebox[0pt][c]{$32 \times$}} %G
\pscircle(-30,-180){10}
\pscircle(10,-180){10}
\pscircle(50,-180){10}
\psline{->}(-20,-180)(-2,-180)
\psline{->}(22,-180)(40,-180)
\pscircle[fillstyle=solid](0,-180){2}
\pscircle[fillstyle=solid](20,-180){2}
\psline{->}(9,-170)(11,-170)
\psline{->}(9,-189)(11,-189)
\rput(-80,-205){\makebox[0pt][c]{$32 \times$}} %H
\pscircle(-30,-205){10}
\pscircle(10,-205){10}
\pscircle(50,-205){10}
\psline{->}(-20,-205)(0,-205)
\psline{<-}(20,-205)(40,-205)
\rput(-80,-230){\makebox[0pt][c]{$32 \times$}} %I
\pscircle(-30,-230){10}
\pscircle(10,-230){10}
\pscircle(50,-230){10}
\psline{<-}(-20,-230)(0,-230)
\psline{->}(20,-230)(40,-230)
\rput(-80,-255){\makebox[0pt][c]{$32 \times$}} %J
\pscircle(-30,-255){10}
\pscircle(10,-255){10}
\psline{->}(-20,-255)(0,-255)
\psline{->}(10,-265)(10,-245)
\psline{->}(19.5,-256)(19.5,-254)
\rput(-80,-280){\makebox[0pt][c]{$64 \times$}} %K
\pscircle(-30,-280){10}
\pscircle(10,-280){10}
\psline{->}(-20,-280)(0,-280)
\psline{->}(10,-290)(10,-270)
\psline{<-}(19.5,-281)(19.5,-279)
\rput(-80,-305){\makebox[0pt][c]{$32 \times$}} %L
\pscircle(-30,-305){10}
\psline{->}(-20,-305)(-2,-305)
\pscircle[fillstyle=solid](0,-305){2}
\pscircle(40,-305){15}
\psline{->}(2,-305)(29.39,-294.39)
\psline{->}(2,-305)(29.39,-315.61)
\rput(-80,-330){\makebox[0pt][c]{$32 \times$}} %M
\pscircle(-30,-330){10}
\pscircle(10,-330){10}
\psline{<-}(-20,-330)(0,-330)
\psline{->}(10,-340)(10,-320)
\psline{->}(19.5,-331)(19.5,-329)
\rput(-80,-355){\makebox[0pt][c]{$64 \times$}} %N
\pscircle(-30,-355){10}
\pscircle(10,-355){10}
\psline{<-}(-20,-355)(0,-355)
\psline{->}(10,-365)(10,-345)
\psline{<-}(19.5,-356)(19.5,-354)
\rput(-80,-380){\makebox[0pt][c]{$32 \times$}} %O
\pscircle(-30,-380){10}
\psline{<-}(-20,-380)(-2,-380)
\pscircle[fillstyle=solid](0,-380){2}
\pscircle(40,-380){15}
\psline{<-}(2,-380)(29.39,-369.39)
\psline{<-}(2,-380)(29.39,-390.61)
\rput(-80,-405){\makebox[0pt][c]{$32 \times$}} %P
\pscircle(-30,-405){12}
\psline{->}(-35,-415)(-35,-395)
\psline{->}(-25,-415)(-25,-395)
\rput(-80,-430){\makebox[0pt][c]{$8 \times$}} %Q
\pscircle(-30,-430){12}
\psline{->}(-35,-440)(-35,-420)
\psline{<-}(-25,-440)(-25,-420)
\psline{->}(-42,-431)(-42,-429)
\rput(-80,-455){\makebox[0pt][c]{$32 \times$}} %R
\pscircle(-30,-455){12}
\psline{->}(-35,-465)(-35,-445)
\psline{<-}(-25,-465)(-25,-445)
\psline{<-}(-42,-456)(-42,-454)
\rput(-80,-495){\makebox[0pt][c]{$64 \times$}} %S
\pscircle(-30,-495){12}
\psline{->}(-30,-505)(-30,-485)
\pscurve{->}(-18,-495)(-14,-480)(-30,-475)(-46,-480)(-42,-495)
\rput(-80,-525){\makebox[0pt][c]{$32 \times$}} %T
\pscircle(-30,-525){12}
\pscircle(10,-525){12}
\psline{->}(-20,-520)(0,-520)
\psline{->}(-20,-530)(0,-530)

%from wikipedia
%\psset{linecolor=black, linewidth=.5pt, arrowsize=2pt 4}
%\newrgbcolor{lightblue}{.8 .82 .86}
%\psset{unit=2.2000cm}
%\pspicture*(-1.6500,-1.6500)(1.7000,1.7000)
%\psset{linewidth=0.1250pt}
%\psset{linecolor=gray}
%\psline(0.0000,0.0000)(0.4157,0.2400)
%\uput{0.1500}[-30.0000](0.0000,1.5500){$y$}
%\uput{0.1500}[-130.0000](1.6000,0.0000){$x$}
%\endpspicture
%
%
%\usepackage{pstricks}
%\usepackage{multido}
%\SpecialCoor
%\pagestyle{empty}
%\parindent=0pt
%\begin{document}
%\begin{pspicture}(-7.5,-7.5)(7.5,7.5)
%   \pscircle{5}
%   \multido{\iA=0+1}{360}{\psline[linewidth=0.1pt](4.6;\iA)(5;\iA)}
%   \multido{\iA=90+-10,\iB=0+10}{36}{\rput{-\iB}(4.3;\iA){\iB}}
%   \multido{\rA=-7.5+2.5}{7}{\psline[linewidth=0.2pt](\rA,-7.5)(\rA,7.5)}
%   \psline[linewidth=0.2pt](-7.5,0)(7.5,0)
%   \pscircle[fillcolor=white,fillstyle=solid]{0.1}
%\end{pspicture}
%http://TeXnik.de/
%http://PSTricks.de/
%ftp://ftp.dante.de/tex-archive/info/math/voss/Voss-Mathmode.pdf
%http://www.dante.de/faq/de-tex-faq/
%http://www.tex.ac.uk/cgi-bin/texfaq2html?introduction=yes

%\vspace*{555pt}
\vspace*{539pt}
\end{center}
\caption{All possible topologically different components with $m=2$, together with their multiplicities (number of ways to construct them). Arcs stand for one link or a chain of nodes with one relevant input. Each crossing of arcs marks a node with two relevant inputs or outputs. Filled points represent exactly one node. Arrows depict the inputs of nodes with two relevant inputs, they appear only where they are important to distinguish the complex component. 
}
\label{fig:m2motifs}
\end{figure}

The same counting of ways to construct complex components for $m=2$
leads to the table presented in \hbox{figure~\ref{fig:m2motifs}}. To
get the corresponding relative frequency of occurrence one has to
divide the multiplicity in figure~\ref{fig:m2motifs} by $6!$. The first
three rows in the figure correspond to distributing nodes with two
relevant inputs or outputs over more than one relevant component,
compare discussion in section \ref{sec:theo}. The dynamical behavior of
the components in figure~\ref{fig:m2motifs} can be studied in details
similarly to \cite{kaufman:on}, if required. It is by analogy clear,
though, that one would get exponentially many exponentially long
attractors with increasing system size.

\section{Conclusions}
\label{sec:final}

In this paper, we have investigated the properties of relevant
components of critical random Boolean networks. We used an efficient
numerical method based on \cite{kaufmanandco:scaling} to create a
sufficiently large ensemble of the relevant parts of sufficiently
large networks in order to evaluate the number of relevant components,
the number of relevant nodes with two relevant inputs, and the number
of components with a given number of nodes with two relevant inputs. The
results are in agreement with theoretical predictions made in
\cite{kaufmanandco:scaling}, and they are supplemented by additional
analytical results. 

Our main findings are the following: 
\begin{itemize}
\item The number of relevant components increases logarithmically with the system size, and usually only the largest relevant component is complex, i.e., is not a simple loop. 
\item For large network sizes, the number of relevant nodes with two relevant inputs and the relative frequencies of different types of complex components become independent of the network size.
\item The relative frequency of topologically different complex components with the same number of nodes with two relevant inputs can be obtained from simple combinatorial considerations. 
\item At constant network size, when the mean number of nonfrozen nodes with two relevant inputs grows (that is, for larger $\kappa=1+\gamma/\beta$), 
 the number of relevant components decreases roughly by the mean value of the logarithm of the number of relevant nodes with two relevant inputs, while the size of the largest relevant component and the number of nodes with two relevant inputs in this component becomes larger.
\end{itemize}

Since the relevant part constitutes only a vanishing portion of the network
(the fraction $N^{-2/3}$ of all nodes), and since the different
relevant components change their state independently of each other, the topology of
considered types of
networks is most likely very different from the topology of real biological networks,
such as genetic regulatory networks, where one would expect that the
majority of nodes is relevant or at least not always frozen, and that
different parts of the network are not decoupled from each
other. Furthermore, we have not addressed the question of the
\emph{function} of a complex component in biological context, which has to be considered together with
a suitable definition of the environment and together with an appropriate choice of update rules.
A lot of work still needs to
be done before real genetic regulatory and other biological networks can be modelled.

\newpage
\bibliography{tpl3}
\end{document}